\documentclass{PoS}

\usepackage{epsf,amssymb,amsthm,amsfonts,amsmath,latexsym, graphicx,array,extarrows,mathtools,mathstyle,mathrsfs,slashed,amsbsy}
\usepackage{cite}

\newcommand{\ud}{\mathrm{d}}
\newcommand{\uvec}[1]{\boldsymbol{#1}}

\title{Gravitational form factor constraints and their universality}

\ShortTitle{}

\author{\speaker{Peter Lowdon} \\
        CPHT, CNRS, Ecole Polytechnique, Institut Polytechnique de Paris, Route de Saclay, 91128 Palaiseau, France \\
        E-mail: \email{peter.lowdon@polytechnique.edu}}

\author{Sabrina Cotogno\\
        CPHT, CNRS, Ecole Polytechnique, Institut Polytechnique de Paris, Route de Saclay, 91128 Palaiseau, France \\
        E-mail: \email{sabrina.cotogno@polytechnique.edu}}
        
        \author{C\'{e}dric Lorc\'{e}\\
        CPHT, CNRS, Ecole Polytechnique, Institut Polytechnique de Paris, Route de Saclay, 91128 Palaiseau, France \\
        E-mail: \email{cedric.lorce@polytechnique.edu}}

\abstract{

By adopting a local QFT framework one can derive in a non-perturbative manner the constraints imposed by Poincar\'e symmetry on the form factors appearing in the Lorentz covariant decomposition of the energy-momentum tensor matrix elements. In particular, this approach enables one to prove that these constraints are in fact independent of the internal properties of the states appearing in the matrix elements. Here we outline the rationale behind this approach, and report on some of the implications of these findings.}

\FullConference{Light Cone 2019 - QCD on the light cone: from hadrons to heavy ions - LC2019\\
		16-20 September 2019\\
		Ecole Polytechnique, Palaiseau, France}

\begin{document}

\section{Introduction}

\noindent
The matrix elements of the energy-momentum tensor (EMT) play an important role in many different physical processes, from gravitational scattering~\cite{Boulware:1974sr,Donoghue:2001qc} to the internal properties of hadrons~\cite{Ji:1994av,Ji:1995sv,Ji:1996ek,Polyakov:2002yz,Goeke:2007fp,Leader:2013jra,Roberts:2016vyn,Lorce:2017wkb,Lorce:2017xzd,Polyakov:2018zvc,Lorce:2018egm,Burkert:2018bqq, Kumericki:2019ddg}. These matrix elements can be decomposed into a series of Lorentz invariant form factors\footnote{These form factors are often referred to as the \textit{gravitational} form factors.}, which fully encode their non-perturbative structure. In recent years, there has been a significant effort, particularly in the context of hadronic physics~\cite{Polyakov:2002yz,Goeke:2007fp,Lorce:2017wkb,Polyakov:2018zvc,Lorce:2018egm,Kumano:2017lhr,Abidin:2008ku,Taneja:2011sy,Cosyn:2019aio,Polyakov:2019lbq}, to try and classify these objects for states of different spin. Here we report on recent progress in this direction~\cite{Cotogno:2019xcl, Lorce:2019sbq}.

\section{The gravitational form factors}

\noindent
In order to analyse the constraints imposed on the gravitational form factors one must first define the physical on-shell states appearing in the EMT matrix elements. For massive states one defines:   
\begin{align}
|p ,\sigma;M \rangle = \delta_{M}^{(+)}(p)|p,\sigma\rangle \equiv 2\pi \,\theta(p^{0})\,\delta(p^{2}-M^{2}) |p,\sigma\rangle,
\end{align}
where $M$ is the mass of the state, $\sigma$ is the canonical spin projection in the $z$-direction, and $|p , \sigma \rangle$ the unrestricted momentum eigenstate. This definition guarantees that the states are manifestly on-shell, and enables the corresponding matrix elements to be analysed in a Lorentz covariant manner. Taking the EMT operator $T^{\mu\nu}$ to be the symmetric version~\cite{Belinfante39}, it follows from the hermiticity and conservation of this current that the matrix elements for \textit{arbitrary} spin can be written\footnote{This decomposition can also in principle be applied to massless states, where now $\sigma$ is the helicity projection of the state. However, non-trivial subtleties arise as the spin of the states is increased. These issues will be addressed in a future work.}~\cite{Boulware:1974sr}
\begin{align}
\langle p',\sigma';M|T^{\mu\nu}(0)|p ,\sigma;M \rangle = \overline{\eta}_{\sigma'}(p')\left[\bar{p}^{\{\mu}\bar{p}^{\nu\}}  A(q^{2}) + i \bar{p}^{\{\mu}S^{\nu\}\rho}q_{\rho} \, G(q^{2}) + \cdots \right]\eta_{\sigma}(p)  \, \delta_{M}^{(+)}(p')\,\delta_{M}^{(+)}(p),
\label{T_decomp}
\end{align}
where $\cdots$ indicates contributions with an explicitly higher-order dependence on the four-momentum transfer $q=p'-p$, and $\bar{p}= \tfrac{1}{2}(p'+p)$ is the average four-momentum. $\eta_{\sigma}(p)$ is the Lorentz-index-carrying coefficient that appears in the canonical decomposition of the free field with the same spin as the state. Collectively, we refer to these coefficients as the \textit{generalised polarisation tensors} (GPTs): for spin-$\tfrac{1}{2}$ these are simply the Dirac spinors $u_{\sigma}(p)$, and for spin-1 they are the polarisation vectors $\varepsilon_{\sigma}(p)$. $S^{\mu\nu}$ are the Lorentz generators in the spin representation of the GPTs. Since the arbitrary spin GPTs appear in a purely external manner in Eq.~\eqref{T_decomp}, this makes it clear that the complexity of this expression, and ultimately the total number of independent form factors, is completely determined by the possible combinations of contracting the Lorentz generators $S^{\mu\nu}$ with $\bar{p}$, $q$, and the metric, whilst respecting the conservation and symmetry of $T^{\mu\nu}$. \\

\noindent
In~\cite{Lowdon:2017idv} it was demonstrated for the first time in the spin-$\tfrac{1}{2}$ case that by expressing the matrix elements of the Poincar\'e charges in two ways: using the form factor expansion, and the Poincar\'e transformation properties of the states, one can equate these equivalent expressions and derive the following zero momentum-transfer constraint on $A(q^{2})$ and $G(q^{2})$:
\begin{align}
A(0)=G(0)=1. \label{FF}
\end{align}
At this point it remained an open question as to whether this constraint continued to hold for massive states of \textit{arbitrary} spin. Using Eq.~\eqref{T_decomp} it was subsequently proven in~\cite{Cotogno:2019xcl} that this is in fact true, and in~\cite{Lorce:2019sbq} this proof was generalised to states defined using \textit{any} spin-state convention. In the remainder of these proceedings we will provide an overview of these results.

\section{Constraints for massive states}

\noindent
As outlined in~\cite{Lowdon:2017idv}, a rigorous operator definition of the rotation $J^{i}$ and boost $K^{i}$ generators is essential for correctly deriving the gravitational form factor constraints\footnote{See~\cite{Lowdon:2017idv} and references within for a discussion regarding the non-perturbative motivation for these definitions.}. Using the expressions for the corresponding currents, these charges are defined:
\begin{align}
&J^{i} = \frac{1}{2}\,\epsilon^{ijk} \lim_{\substack{d \rightarrow 0 \\ R \rightarrow \infty}}\int \ud^{4}x \, f_{d,R}(x) \left[ x^{j}T^{0k}(x) - x^{k}T^{0j}(x) \right], \label{j_charge}\\
&K^{i} =  \lim_{\substack{d \rightarrow 0 \\ R \rightarrow \infty}}\int \ud^{4}x \, f_{d,R}(x) \left[ x^{0}T^{0i}(x) - x^{i}T^{00}(x) \right], \label{k_charge}
\end{align}
where $f_{d,R}(x) \equiv \alpha_{d}(x^{0})F_{R}(\uvec{x})$, and the (test) functions $\alpha_{d}$ and $F_{R}$ satisfy the conditions: $\int \ud x^{0}  \alpha_{d}(x^{0}) =1$, $\lim_{d\rightarrow 0}\alpha_{d}(x^{0}) = \delta(x^{0})$, $F_{R}(\uvec 0) = 1$, and: $\lim_{R\rightarrow \infty}F_{R}(\uvec{x}) = 1$. Combining these definitions with the form factor decomposition in Eq.~\eqref{T_decomp}, one obtains:
\begin{align}
\langle p',\sigma';M|J^{i}|p ,\sigma ;M\rangle &= -i\epsilon^{ijk}\bar{p}^{k}(2\pi)^{4}\delta_{M}^{(+)}(\bar p) \, A(q^2)\left[ \delta_{\sigma'\sigma} \, \partial^{j} - \left.\partial^{j} \!\left[\overline{\eta}_{\sigma'}(p')\eta_{\sigma}(p)\right]\right|_{q=0} \right] \delta^4(q) \nonumber \\
&\quad \quad +\frac{1}{2}\,\epsilon^{ijk} (2\pi)^{4}\delta_{M}^{(+)}(\bar p) \,  G(q^2)  \left[ \overline{\eta}_{\sigma'}(\bar{p})S^{jk}\eta_{\sigma}(\bar{p}) \right] \delta^4(q)  \label{j_m}\\
\langle p',\sigma';M|K^{i}|p ,\sigma ;M\rangle &= i (2\pi)^{4}\delta_{M}^{(+)}(\bar p) \, A(q^{2})\left[\delta_{\sigma'\sigma}\,(\bar{p}^{0} \partial^{i} - \bar{p}^{i} \partial^{0})- \left.\bar{p}^{0}\partial^{i}\!\left[\overline{\eta}_{\sigma'}(p')\eta_{\sigma}(p)\right]\right|_{q=0}      \right]\delta^4(q)  \nonumber \\
& \quad \quad + (2\pi)^{4}\delta_{M}^{(+)}(\bar p) \,  G(q^{2})\left[\overline{\eta}_{\sigma'}(\bar{p})S^{0i}\eta_{\sigma}(\bar{p})  \right] \delta^4(q)   \label{k_m}.
\end{align}
One can also write an alternative representation of these matrix elements using the general Lorentz transformation condition for massive (canonical spin) states of spin $s$~\cite{Haag:1992hx}:
\begin{align}
U(\alpha)|p,\sigma;M\rangle = \sum_{\sigma'}\mathcal{D}^{(s)}_{\sigma'\sigma}(\alpha)|\Lambda(\alpha)p,\sigma';M\rangle,
\label{Lorentz_tran}
\end{align}
where $\mathcal{D}^{(s)}(\alpha)$ is the $(2s+1)$-dimensional Wigner rotation matrix corresponding to the (proper orthochronous) Lorentz transformation $\alpha$, and $\Lambda(\alpha)$ is the four-vector representation of $\alpha$. By separately expanding this expression for infinitesimal rotations and boosts, and using the distributional characteristics of the states, the $J^{i}$ and $K^{i}$ matrix elements have the following general structure:
\begin{align}
\langle p',\sigma';M| J^{i}|p,\sigma;M\rangle &= (2\pi)^{4}\delta_{M}^{(+)}(\bar{p})\left[\Sigma^{i}_{\sigma'\sigma}(k) - \delta_{\sigma'\sigma} \, i\epsilon^{ijk}\bar{p}^{k}\partial^{j} \right]\delta^{4}(q), \label{J_M}  \\
\langle p',\sigma';M| K^{i}|p,\sigma;M\rangle &= (2\pi)^{4}\delta_{M}^{(+)}(\bar{p})\bigg[-\frac{\epsilon^{ijk}\bar{p}^{j}}{\bar{p}^{0}+M}\, \Sigma^{k}_{\sigma'\sigma}(k) + \delta_{\sigma'\sigma} \, i \left(\bar{p}^{0} \partial^{i} - \bar{p}^{i} \partial^{0} \right) \bigg]\delta^{4}(q),
\label{K_M} 
\end{align}
where $\Sigma^{i}_{\sigma'\sigma}(k) = \overline{\eta}_{\sigma'}(k)J^{i}\eta_{\sigma}(k)$ is the rest-frame spin of the state, and $k = (M,0,0,0)$ the (canonical) rest-frame vector. After explicitly performing the derivatives in the GPT terms in Eqs.~\eqref{j_m} and~\eqref{k_m}, and equating these expressions with Eqs.~\eqref{J_M} and~\eqref{K_M} respectively, this implies in both cases that the form factors must satisfy the following constraints:
\begin{align}
A(q^{2})\,\delta^4(q) &= \delta^4(q), \label{constr1} \\
A(q^{2})\,\partial^{j}\delta^4(q) &= \partial^{j}\delta^4(q), \label{constr2} \\
G(q^{2})\,\delta^4(q) &= \delta^4(q), \label{constr3}
\end{align}
which is simply the condition in Eq.~\eqref{FF}. This proves that the zero-momentum transfer constraint on the gravitational form factors is completely independent of the spin of the states appearing in the EMT matrix elements. It turns out that one can \textit{also} repeat the same procedure for the covariantised Lorentz generators, the \textit{Pauli-Lubanski} operator: $W^{\mu}= \tfrac{1}{2}\epsilon^{\mu}_{\phantom{\mu} \rho\sigma\lambda}M^{\rho\sigma}P^{\lambda}$ and the \textit{covariant boost}: $B^{\mu} = \tfrac{1}{2}\left[ S^{\nu\mu}P_{\nu} + P_{\nu}S^{\nu\mu} \right]$. Doing so, one finds that the $W^{\mu}$ and $B^{\mu}$ matrix elements separately imply the constraints $G(0)=1$ and $A(0)=1$, respectively. In other words, using the covariantised operator basis results in a diagonalisation of the constraints. Since the only dynamical condition that enters these calculations is the transformation properties of the states [Eq.~\eqref{Lorentz_tran}], these results imply that the Poincar\'e symmetry of the theory alone is responsible for determining the behaviour of $A(q^{2})$ and $G(q^{2})$ as $q \rightarrow 0$.

\section{Arbitrary state generalisation}
\noindent
As with any relativistic spin states, the massive canonical spin states discussed in the previous section are convention dependent, and defined via the action of a specific boost on a rest-frame state. The specific form of this boost is encoded in the GPT derivative terms in Eqs.~\eqref{j_m} and~\eqref{k_m}, and plays an essential role in the form factor constraint calculations in~\cite{Cotogno:2019xcl}. It therefore remains unclear from these results alone as to whether these constraints are actually dependent upon the choice of spin-state convention. In general, spin states $|p ,\sigma \rangle$ are defined~\cite{Haag:1992hx}:
\begin{align}
|p , \sigma\rangle = U(L(p))|k,\sigma\rangle,
\label{spin_def}
\end{align}
where $L(p)$ is some choice of Lorentz transformation which maps a reference frame four-vector $k$ to an arbitrary (on-shell) four-momentum $p$
\begin{align}
\Lambda(L(p))k = p.
\end{align}
For massive states, $\sigma$ labels the rest frame spin projection along some axis, whereas for massless states $\sigma$ corresponds to the helicity projection of the state along the direction of motion. Besides canonical spin states, where $k=(M,0,0,0)$ and $L_{\text{c}}(p)$ is a pure boost along the direction $\hat{\uvec{p}} = \tfrac{\uvec{p}}{|\uvec{p}|}$, important spin state examples include~\cite{Bakker:2004ib,Polyzou:2012ut}: (\textit{i}) \textit{Wick helicity states} -- $L_{\text{W}}(p)$ is a pure boost along the $z$-direction followed by a rotation into $\hat{\uvec{p}}$; (\textit{ii}) \textit{Light-front spin states} -- $L_{\text{LF}}(p)$ is a pure boost along the $z$-direction followed by a transverse light-front boost. Unlike canonical spin states, Wick helicity and light-front spin also make sense for \textit{massless} states, where instead $k= (\kappa,0,0,\kappa)$ with $\kappa >0$. Since the EMT matrix elements are parametrised purely in terms of GPTs, in order to generalise the previous analysis to arbitrary choices of spin states, one must understand how the GPTs themselves depend on the choice of $L(p)$ and $k$. In general, one finds that~\cite{Lorce:2019sbq}:
\begin{align}
\eta_{\sigma}(p) &= D(L(p))\eta_{\sigma}(k), \label{eta_rel} 
\end{align}   
where $D$ is the finite-dimensional Lorentz representation of the (free) field of which $\eta_{\sigma}(p)$ is a component\footnote{In other words, $D$ is the finite-dimensional matrix appearing in the Lorentz transformation of the field $\varphi$ associated with the GPT: $U(\alpha)\varphi_{i}(x)U(\alpha)^{-1} = \sum_{j}D_{ij}(\alpha^{-1})\varphi_{j}(\Lambda(\alpha)x)$.}. Applying these definitions, one finds that the $J^{i}$ matrix elements can be written:
\begin{align}
\langle p',\sigma';M|J^{i}|p ,\sigma ;M\rangle = &-i\epsilon^{ijk}\bar{p}^{k} \delta_{\sigma'\sigma}(2\pi)^{4}\delta_{M}^{(+)}(\bar p) \, \partial^{j}\delta^4(q) A(q^2) \nonumber \\ &+i\epsilon^{ijk}\bar{p}^{k} (2\pi)^{4}\delta_{M}^{(+)}(\bar p) \,  \overline{\eta}_{\sigma'}(k)\widetilde{D} \! \left( \frac{\partial L^{-1}(\bar{\uvec{p}})}{\partial \bar{p}_{j}} L(\bar{\uvec{p}}) \right)\eta_{\sigma}(k)  \,  \delta^4(q)  A(q^2) \nonumber \\
& + (2\pi)^{4}\delta_{M}^{(+)}(\bar p) \overline{\eta}_{\sigma'}(k)\widetilde{D}\left(L^{-1}(\bar{\uvec{p}})J^{i}L(\bar{\uvec{p}})\right)\eta_{\sigma}(k) \, \delta^4(q) \,  G(q^2),
\label{J_L}
\end{align}
where $\widetilde{D}$ is the Lie algebra representation of $D$. Similarly, one can also re-write Eq.~\eqref{J_M} explicitly in terms of $L(p)$, which upon comparison with Eq.~\eqref{J_L} implies the conditions in Eqs.~\eqref{constr1},~\eqref{constr2} and~\eqref{constr3}. The same is also true for the boost matrix elements. Since the form of $L(p)$ is unspecified in Eq.~\eqref{J_L}, this proves that the form factor constraint $A(0)=G(0)=1$ is \textit{completely independent} of the internal characteristics of the states in the EMT matrix elements, and that Poincar\'{e} symmetry alone is responsible for this condition.

\section{Implications and outlook}
\noindent
Although the zero-momentum transfer limit of the gravitational form factors had been discussed many times over the years for massive states of lower spin, until recently it remained an open question as to whether this limit continued to hold for states with higher spin, or for different choices of definition of the spin states themselves. As outlined in these proceedings, these questions were definitively addressed in~\cite{Cotogno:2019xcl} and~\cite{Lorce:2019sbq}, where is was proven for the first time that the constraint $A(0)=G(0)=1$ is true for states of both arbitrary spin and spin-state convention. These findings have many potential implications. For example, in the context of hadronic physics this result implies that the \textit{Ji sum rules} satisfied by generalised parton distributions (GPDs)~\cite{Ji:1996ek} are in fact completely independent of the choice of states used in the definition of the GPDs, and are therefore not specific to composite spin-$\tfrac{1}{2}$ states such as the proton. From a gravitational perspective, this condition implies that any particle moving in an external gravitational field must necessarily have a vanishing anomalous gravitomagnetic moment~\cite{Teryaev:1999su}. Overall, the general framework developed in~\cite{Cotogno:2019xcl} and~\cite{Lorce:2019sbq} opens a new direction for exploring the non-perturbative structure of EMT matrix elements, with the potential to provide new insights into other interesting questions, such as determining the number of independent form factors that exist for states of a given spin, or whether additional constraints arise in theories with more symmetries, such as conformal field theories. The first of these questions has in fact been addressed in a recent work~\cite{Cotogno:2019vjb}.

\section*{Acknowledgements}
\noindent
This work was supported by the Agence Nationale de la Recherche under the Projects No. ANR-18-ERC1-0002 and No. ANR-16-CE31-0019.

\bibliographystyle{JHEP}

\bibliography{paper_refs}

\end{document}